\documentclass[aps,prc,groupedaddress,showpacs,showkeys]{revtex4}
\usepackage{graphicx}
\begin{document}

%\preprint{}

\title{Behavior of shell-model configuration moments}

\author{Edgar Ter\'an}
%\email[]{terane@sciences.sdsu.edu}
%\homepage[]{http://www.rohan.sdsu.edu/~terane/}
\author{Calvin W. Johnson}
%\email[]{cjohnson@sciences.sdsu.edu}
%\homepage[]{http://www.physics.sdsu.edu/~johnson/}
%\thanks{}
\affiliation{Department of Physics, San Diego State University \\
5500 Campanile Dr, San Diego, CA 92182-1233, USA}


\date{\today}

\begin{abstract}
An important input into reaction theory is the density of states or
the level density. Spectral distribution theory (also known as
nuclear statistical spectroscopy) characterizes the secular behavior
of the density of states through moments of the Hamiltonian. One
particular approach is to partition the model space into subspaces
and find the moments in those subspaces; a convenient choice of
subspaces are spherical shell-model configurations. We revisit these
configuration moments and find, for complete $0\hbar\omega$ many-body spaces, 
the following behaviors: (a) the
configuration width is nearly constant for all configurations; (b)
the configuration asymmetry or third moment is strongly correlated
with the configuration centroid; (c) the configuration fourth
moment, or excess  is linearly related to the square to the
configuration asymmetry.  Such universal behavior may allow for more
efficient modeling of the density of states in a shell-model
framework.

\end{abstract}

\pacs{21zc}

\maketitle

\section{Introduction and Motivation}

An important input into Hauser-Feshbach calculations of statistical
capture of neutrons \cite{HaFe52,RaThKr97} is the density of nuclear states, or
the nuclear level density, as a function of excitation energy. (The phrases
``state density'' and ``level density'' are often used informally as
synonyms, although they are not: the state density includes $2J+1$
degeneracies, where $J$ is the angular momentum. Because the
Hauser-Feshbach formalism uses $J$-dependent transmission
coefficients, one ultimately requires the level density as a
function of $J$. As will become clear below, we discuss both the
total state density as well as $J$-projected level densities.)

The density of states is difficult to obtain both experimentally and
theoretically and is perhaps the most uncertain input into
Hauser-Feshbach calculations \cite{RaThKr97}.  Therefore there is
significant effort, both experimental and theoretical, going into
determining the density of states.

The most common theoretical approaches are built around
noninteracting Fermi gases, starting from Bethe's formula\cite{Be36}
using the single-particle density of
states. Of course the residual interaction plays a significant role,
and  single-particle approaches necessarily add corrections to
account for the residual interaction\cite{Go96,DeGo01}.

An alternative is to start from a fully interacting picture such as
the configuration-mixing shell model. Interacting shell-model codes,
such as OXBASH\cite{OXBASH}, ANTOINE\cite{ANTOINE}, and
REDSTICK\cite{Redstick},  read in single-particle
energies and two-body matrix elements and compute matrix elements of
the many-body Hamiltonian in a truncated, albeit very large,
many-body basis. All such codes use the Lanczos algorithm to extract
the low-lying spectra\cite{Wh77}. Obtaining the complete spectrum
is unfeasible for large dimension bases, and unnecessary as well:
one only needs the smooth, secular behavior of the density of
states, not the individual states. Instead, one turns to
formulations of the interacting shell model that do not involve
diagonalization: auxiliary-field path integrals \cite{Or97,NaAl97,NaAl98},
or spectral distribution theory (also known as nuclear statistical
spectroscopy\cite{FrRa71,AyGi74,MoFr75,Wo86}).

Nuclear statistical spectroscopy starts from the moments of the
Hamiltonian. For example, it is known that, in a finite space, most
state densities of many-body systems with a two-body interaction
tend toward a Gaussian shape \cite{MoFr75,Ko84} 
characterized by the first and
second moments of the Hamiltonian. In most realistic cases the
density has small but non-trivial deviations from a Gaussian, so one
requires higher moments. It has long been realized that, rather than
computing higher and higher moments of the total Hamiltonian, one
could partition the model space into suitable subspaces and compute
just a few moments in each subspace. In particular, if one
partitions the space based upon spherical shell-model
configurations, that is, all states of the form
$(0d_{5/2})^4(1s_{1/2})^2 (0d_{3/2})^2$, etc., then it is possible
to derive expressions for the configuration moments directly in
terms of the single-particle energies and two-body matrix elements,
without constructing any many-body matrix elements (see, however,
our discussion in depth below).

Although such an approach has obvious merit, it is far from trivial.
For a realistic application, one must include thousands 
of configurations. For example, $^{54}$Fe in a
$1s0d$-$1p0f$ space allowing at most two particles to be excited out
of the $sd$-shell into the $pf$ consists of 147,060 configurations
(which in turn represent $9.5 \times 10^{11}$ levels). Furthermore,
higher moments take more and more computational effort. The
configuration centroids and widths (first and second moments,
respectively) take CPU-minutes, while configuration
asymmetries (third moments) take CPU-hours or days and configuration 4th
moments even longer; in our example of $^{54}$Fe it takes about 43
minutes on a single 2.4 GHz Pentium to compute all the second
moments, but 9 days to compute the third moments.

Therefore one important question is: how many moments must be
reliably calculated? A related question is: what is an appropriate
model for the secular behavior? Most work has been based on
modified Gaussians, in the Gram-Charlier or Edgeware distributions. 
Kota, Potbhare, and Shenoy\cite{KoPoSh86} made a study of
model functions and
concluded the Cornish-Fisher distribution as the best approximation;
they also made a study of the behavior of the configuration moments,
and this paper can be considered an further investigation of that
early work. Zuker\cite{Zu01} proposed binomial distributions as a 
way to naturally include asymmetries. Recently Horoi, Kaiser, and Zelevinsky
\cite{HoKaZe03} have modeled the level density as a
sum of Gaussians, based only upon first and second moments, and
ignoring higher moments such as the asymmetry; they argue that the
asymmetries tend to cancel.

Given this background, we have investigated in more detail the configuration
moments and their distributions. Before sharpening our questions further,
we lay out more carefully the formalism of spectral distribution theory
and configuration moments.

\section{Formalism and more detailed motivation}

We begin with the moments of a many-body Hamiltonian, taking some
care to define our formalism; this is important because we will
investigate if the behavior of moments depends upon, for example,
angular momentum. None of the formalism in this section 
is original; a thorough reference to statistical 
spectroscopy is Ref.~\cite{Wo86}, although some of our notation is different.

 We always work in a finite model space
${\cal M}$ wherein the number of particles is fixed. If in ${\cal
M}$ we represent the Hamiltonian as a matrix $\mathbf{H}$, then all
the moments can be written in terms of traces; for some matrix
$\mathbf{O}$ the trace is
\begin{equation}
\mathrm{tr\,} \mathbf{O} = \sum_{i \in {\cal M}}
\langle i | \mathbf{O} |i \rangle
\end{equation}
The total dimension of the space is $D = \mathrm{tr\,}\mathbf{1}$, and then the
average is
\begin{equation}
\langle \mathbf{O} \rangle = \frac{1}{D} \mathrm{tr \,} \mathbf{O}.
\end{equation}
The first moment, or the centroid, of the Hamiltonian is then
\begin{equation}
\bar{E} = \langle \mathbf{H} \rangle;
\end{equation}
all other moments are \textit{central} moments, computed relative to
the centroid:
\begin{equation}
\label{eq:shell_moms}
\mu^{(n)} = \langle (\mathbf{H}-\bar{E})^n \rangle \; \; .
\end{equation}

The moments can also be computed directly from the density of states
\begin{eqnarray}
\rho(E) = \mathrm{tr \,} \delta(E-\mathbf{H}), \\
\bar{E} = D^{-1} \int dE \, E \rho(E), \\
\mu^{(n)}= D^{-1} \int dE \, (E-\bar{E})^n \rho(E).
\end{eqnarray}

The width $\sigma$ is given by $\sqrt{\mu^{(2)}}$, and one typically scales
the higher moments by the width:
\begin{equation}
m^{(n)} = \frac{\mu^{(n)}}{\sigma^n}.
\end{equation}

In addition to the centroid and the width, the next two moments have
special names. The scaled third moment $m^{(3)}$ is the
\textit{asymmetry}, or the skewness, while $m^{(4)}-3$ is the
\textit{excess}, so called because for a Gaussian $m^{(4)}=3$ and
hence a Gaussian has zero excess.

In all of the above, one has to be careful to clearly state the model
space ${\cal M}$. If, for example, ${\cal M}$ contains only states of a
fixed $J_z$, then one has the level density. On the other hand, if
${\cal M}$ contains all $J_z$ values, so that one has all $2J+1$ degeneracies,
one has the state density.

Next we consider projection operators and subspaces. For this paper
we use subspaces based upons spherical single-particle
configurations, that is, sets of states of the form, e.g.,
$(0d_{5/2})^4 (0d_{3/2})^2$,  $(0d_{5/2})^4 (0d_{3/2})^1
(1s_{1/2})^1$, etc.. Angular momentum alsos play an
important role, so we adopt nomenclature as follows:

 A $J$-summed configuration means a sum over
all  $J$ and all $J_z$ in
the configuration; $J$-projected means a sum just over states of 
fixed $J$ in a configuration. We can
relate the $J$-summed density, which is really the state density, to
a sum over $J$-projected (level) densities:
\begin{equation}
\rho(E) = \sum_J (2J+1) \rho_{J}(E).
\label{jsum}
\end{equation}

One can in principle also project out states of good isospin. 
For simplicity we consider only $pn$-configurations, 
that is, configurations that have the proton and
neutron occupations fixed separately, i.e., $(0d_5/2)^2_\pi (0d_3/2)^2_\nu$ is
distinguished from $(0d_5/2)^2_\nu (0d_3/2)^2_\pi$ (which, for an
isospin-conserving interaction, will have the same moments) and from
$(0d_5/2)^1_\pi (0d_5/2)^1_\nu (0d_3/2)^1_\pi (0d_3/2)^1_\nu $
(which, for an interaction that does not mix isospin but does depend
on isospin, will generally have different moments).  
We do not believe that our results will differ dramatically for 
projection onto isospin.

Finally we turn to the key question of subspaces.  We use $\alpha,
\beta, \gamma, \ldots$ to label subspaces. Let
\begin{equation}
P_\alpha =\sum_{i \in \alpha} \left | i \right \rangle \left \langle
i \right |
\end{equation}
be the projection operator for the $\alpha$-th subspace.
One can introduce \textit{partial} or \textit{configuration density},
\begin{equation}
\rho_\alpha(E) = \mathrm{tr} \, P_\alpha \delta(E-\mathbf{H} ).
\end{equation}
The $J$-summed and $J$-projected configuration densities are defined
in the obvious way (and because projection of angular momentum
commutes with projection onto single-particle configurations, leads to
no difficulties), and can be related by $\rho_{\alpha}(E) = \sum_J (2J+1)
\rho_{\alpha,J}(E).$

Now we can
define \textit{configuration moments}: the configuration dimension is
$D_\alpha = {\rm tr \,} P_\alpha$, the configuration centroid is
\begin{equation}
\bar{E}_\alpha = D_\alpha^{-1} {\rm tr \,} P_\alpha \mathbf{H},
\end{equation}
while the configuration width $\sigma_\alpha$ is given by 
\begin{equation}
\sigma_\alpha^2 = D_\alpha^{-1} {\rm tr \,} P_\alpha
(\mathbf{H}-\bar{E}_\alpha)^2,
\end{equation}
the configuration asymmetry is
\begin{equation}
m^{(3)}_\alpha = \left(D_\alpha^{-1} {\rm tr \,} P_\alpha
(\mathbf{H}-\bar{E}_\alpha)^3)\right) /\sigma_\alpha^{3},
\end{equation}
and the configuration excess is defined similarly.

Note that in all the above traces, one inserts only one projection operator.
This means
that, although one is working with the ``density'' in a particular subspace,
matrix
elements of the Hamiltonian can take one out of the subspace. For example, one
can further
define, distinct from the configuration width, the \textit{partial variance}
(sometimes called, misleadingly, the partial width)
\begin{equation}
\Gamma_{\alpha \beta} \equiv \frac{1}{D_{\beta} - \delta_{\alpha
\beta}}\left [ \frac{1}{D_\alpha} {\mathrm tr \,} \left( P_\alpha
{\mathbf H} P_\beta {\mathbf H} \right ) - \delta_{\alpha \beta}
\bar{E}_\alpha^2 \right ]
\end{equation}
Then the configuration variance $\sigma_\alpha^2$ arises
from two contributions, an intrinsic piece $\Gamma_{\alpha \alpha}$ and a sum 
over extrinsic pieces $\Gamma_{\alpha \beta}$. Although
some authors make extensive use of partial widths, we will not use them 
in this paper except to understand the origin of correlations in
the configuration asymmetries.

\subsection{Monopole and traceless interactions}

One useful result from spectral distribution theory is that the
configuration centroids depend entirely upon the single-particle
energies and the monopole-monopole part of the residual
interaction\cite{DuZu99}.
The monopole interaction is attributed to mean-field and saturation
properties of the nuclear interaction. One can subtract out the
monopole interaction exactly, which sets the centroids to zero but
leaves the widths unchanged; such a monopole-subtracted interaction
is referred to as a ``traceless'' interaction. Third and fourth
configuration moments \textit{are} changed by subtracting out the
monopole interaction, as we will see below.

We briefly review calculation of the centroids because they give the
nonexpert a better idea of the methods of spectral distribution
theory. We label spherical orbits such as $0d_{5/2}$, $1s_{1/2}$,
etc, by $a,b,c,\ldots$. Let $\epsilon_a$ be the single-particle
energy for orbit $a$ and let  $| ab, J \rangle$ be an
antisymmeterized, normalized 2-particle state with angular momentum
$J$ (it turns out that the label $J_z$ is unnecessary). Then let
$V_J(ab,cd)$ be the matrix elements of the Hamiltonian between
$|ab,J\rangle$ and $ |cd,J\rangle$, after subtracting off the
single-particle energies. The \textit{monopole potential} is then
given by
\begin{equation}
U(ab) = \frac{1+\delta_{ab}}{N_a (N_b+\delta_{ab})} \sum_J
(2J+1)V_J(ab,ab),
\label{monopole}
\end{equation}
where $N_a = 2j_a +1$ is the degeneracy of the orbit $a$.

Any spherical configuration $\alpha$ is defined by the occupation of
the spherical orbits $n_1, n_2, n_3, \ldots$. Then the ($J$-summed)
configuration centroid is
\begin{equation}
\bar{E}_\alpha = \sum_a n_a \epsilon_a + \frac{1}{2} \sum_{ab} n_a
(n_b - \delta_{ab}) U(ab).
\end{equation}
A \textit{traceless} interaction is made by subtracting off the
monopole potential, $\Delta V_J(ab,cd) =
V_J(ab,cd)-\delta_{ac}\delta_{bd}U(ab)$. One uses the traceless
interaction to compute the configuration width, although the
resulting formula is now fourth-order in the $n_a$.

One can extend the above formulas to isospin, although we find it
convenient to work in a $pn$ formalism where protons and neutrons
are assigned to distinct orbits.

\section{Some questions}
\label{questions}

Now that we have reviewed the required mathematical language we can
 ask some interesting questions.  For example:

\textbf{Q1}: \textit{What is the distribution of the configuration
widths?} Do they depend upon the dimension of the configuration?
Upon $A$? We find: the configuration widths are remarkably constant
for a given nuclide, but depend upon $A$ and $N-Z$.

\textbf{Q2}: \textit{What is the distribution of the configuration
asymmetries?}  What range do the configuration asymmetries have? To
model the partial (configuration) level densities it may be useful
to know the range of the configuration asymmetries
$m^{(3)}_{\alpha}$, since many numerical models often have
restrictions in this respect; for example the binomial distribution\cite{Zu01}
has difficulty with $|m^{(3)}| > 1$ \cite{TeJo05}.

Are the asymmetries correlated with, for example, the configuration
centroid? This is relevant because Horoi, Kaiser, and Zelevinsky\cite{HoKaZe03} 
speculate that the
asymmetries in configuration densities tend to cancel out. This
speculation is ill-founded, because in fact there is a strong
correlation of the configuration asymmetry with the the
configuration centroid.  In particular, configurations low in energy 
often, but not always, 
have a significant positive asymmetry (skewed towards higher
energies) while configurations at high energy have significant
negative asymmetry. (Such a correlation can be seen in Fig.~1 of
Ref.~\cite{KoPoSh86}, although those authors did not comment on it.) This
means, at any given excitation energy, all of the ``nearby''
configurations that contribute to the level density have nearly the
same asymmetry.  We note, however, that Horoi \textit{et al}\cite{HoKaZe03} claim satisfactory
modeling of the level density without asymmetries; while this 
result requires further investigation, we speculate they may have happened 
upon fortuitous cases where the asymmetry at low energy is small.

The correlation between the asymmetries and the centroids is
intriguing.  We suggest that a large asymmetry arises through
connections to distant configurations. One can divide the
configuration third moment $m_\alpha^{(3)}$ into two parts, an intrinsic piece
which
comes strictly from matrix elements that do not take you out of the
subspace, and a part that looks like
\begin{equation}
\sum_\beta  D_\beta \Gamma_{\alpha \beta}(\bar{E}_\beta -
\bar{E}_\alpha). \label{asym_cent}
\end{equation}
Such terms could explain the correlation between the configuration
asymmetries and the configuration centroids. We test this idea by
setting all the centroids $\bar{E}_\alpha = 0$, that is, by
considering so-called ``traceless'' interactions. When we do this,
we find the configuration asymmetries are dramatically reduced. 
Furthermore, deviations from this trend we ascribe to collectivity.

\textbf{Q3}: \textit{What is the distribution of the excess?} In
particular we investigate correlations between the excess (fourth
moment) and asymmetry (third moment). If there were a correlation,
it would help in modeling the state density, both in the choice of
appropriate model functions for the secular behavior, but also save
computations: one could compute only the asymmetry and estimate the
excess. We find a strong correlation, that in fact the excess is
linearly related to the square of the asymmetry, with high
statistical significance. The exact parameters of the linear
correlation depend weakly upon the interaction and the nuclide in
question.

\section{Details of calculation methods}

We work in a spherical shell-model basis, and the input to our
calculations are the same as that of most spherical $M$-scheme
codes: a list of single-particle
orbits, and a list of single-particle energies and two-body matrix
elements. In particular we work in the
$1s_{1/2}$-$0d_{3/2}$-$0d_{5/2}$ or \textit{sd} space, where we use
the Brown-Wildenthal universal \textit{sd} (USD)
interaction\cite{Wi84}, and in the
$1p_{1/2}$-$1p_{3/2}$-$0f_{5/2}$-$0f_{7/2}$ or \textit{pf} shell,
where we use the interactions GXPF1 \cite{HoOtBrMi04}, FPD6G
\cite{RMJB,PoSaCaNo01}, and KB3G\cite{Kubr68,PoZu81,PoSaCaNo01}. 
We also used quadrupole-quadrupole ($Q\cdot
Q$) interaction and Gaussian-distributed random two-body
interactions (the two-body random ensemble or TBRE) in both spaces to
investigate any dependence of our
results on the interaction. Finally, we also create ``traceless''
interactions, by subtracting off the monopole potential,  
Eqn.~(\ref{monopole}), so that the centroids $E_\alpha = 0$ identically.

In principle, one can compute $J$-summed, $pn$-configuration moments
of order one through four, starting with single-particle energies
and two-body matrix elements, without constructing the many-body
matrix elements\cite{FrRa71,AyGi74,Wo86}. We have found the formulas
for first and second moments work as stated, but however we were
unable to satisfactorily validate the trace formulas for third and
fourth configuration moments. (At this time we are still unsure if
the fault lies in the published formulas or in our implementation of
them.)

Therefore for this work we used the REDSTICK shell-model
code\cite{Redstick} (which uses input files similar to those 
for OXBASH\cite{OXBASH}) to construct the many-body matrix elements and to
compute the configuration moments directly. This is inefficient, and
clearly for large-scale computations one would need to
satisfactorily implement trace formulas for all required moments. On
the other hand, our direct method allows us to exactly compute
$J$-projected moments as well, something which cannot be done
exactly by trace formulas, although approximate methods do
exist\cite{JaSp79,VeBr82,Wo86}.

For this paper we consider only complete $0\hbar\omega$ many-body spaces, 
due to our current computational limitations. It will be interesting in future
work to consider multi-shell spaces.

\section{\label{moments} Results}

\subsection{Distribution of configuration widths}

\begin{figure}
\includegraphics[scale=0.65]{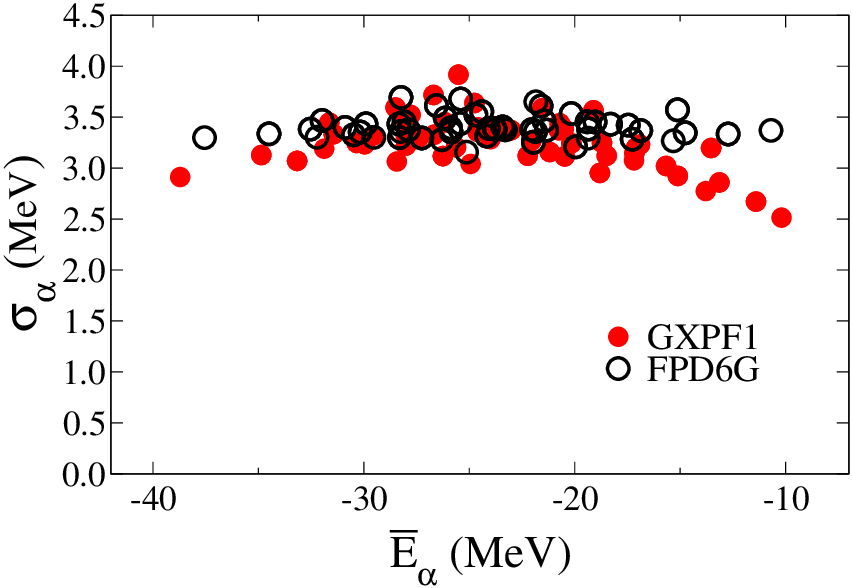}
\caption{\label{widthdist} (Color online) Distribution of configuration in
$^{44}$Ti as function of the corresponding configurations centroids, in the
{\it
pf}-shell with interactions GXPF1 \cite{HoOtBrMi04} and FPD6G
\cite{RMJB,PoSaCaNo01}}
\end{figure}

We begin by considering the distribution of configuration widths. It
is already known that the configuration widths are approximately
constant\cite{Wo86}, which we illustrate in
Fig.\ref{widthdist} for $^{44}$Ti in the $pf$-shell (2 valence
protons and 2 valence neutrons), using all 100 $J$-summed $pn$-configurations.
We consider two interactions, GXPF1 (filled circles) and FPD6G (open
circles), which, unsurprisingly, yield similar results. The
spreading of the widths corresponding to the FPD6G interaction is
very narrow. The GXPF1 widths show a slightly greater range.  

While the configuration widths are constant for a given nuclide, they do 
depend on the nuclide in question. 
Table~\ref{table:widths} shows the average configuration widths (and
corresponding standard deviations) for a number of representative nuclides 
with valence particles in the $sd$-shell with the USD interaction \cite{Wi84}.
The widths are largest for ``half-filled'' cases and for $N \sim Z$ nuclides. 
The scatter, as represented by the standard deviations, is nearly constant as
well, except at nearly full or nearly empty spaces. 

\begin{table}[htb]
\caption{\label{table:widths}
Average configuration widths and corresponding standard deviations for several
nuclides in the {\it sd}-shell with USD \cite{Wi84} interaction.}
\begin{ruledtabular}
\begin{tabular}{ l c c}
Nuclide              & $\langle  \sigma_\alpha \rangle$ (MeV)   & $\pm \Delta
\sigma$ (MeV)  \\
\hline
$^{28}$Ne   &  4.84  & 1.17  \\
$^{28}$Na   &  6.68  & 0.84  \\
$^{28}$Mg   &  8.10  & 0.76  \\
$^{28}$Al   &  8.94  & 0.71  \\
$^{22}$Si   &  3.88  & 0.74  \\
$^{23}$Si   &  5.86  & 0.69  \\
$^{24}$Si   &  7.22  & 0.73  \\
$^{25}$Si   &  8.18  & 0.70  \\
$^{26}$Si   &  8.84  & 0.71  \\
$^{27}$Si   &  9.18  & 0.70  \\
$^{28}$Si   &  9.24  & 0.70  \\
$^{29}$Si   &  8.99  & 0.69  \\
$^{30}$Si   &  8.47  & 0.68  \\
$^{31}$Si   &  7.66  & 0.66  \\
$^{32}$Si   &  6.62  & 0.67  \\
$^{33}$Si   &  5.25  & 0.62  \\
$^{34}$Si   &  3.40  & 0.65  \\
\end{tabular}
\end{ruledtabular}
\end{table}

\subsection{Distribution of  configuration asymmetries}

\begin{figure}[t]
\includegraphics[scale=0.65]{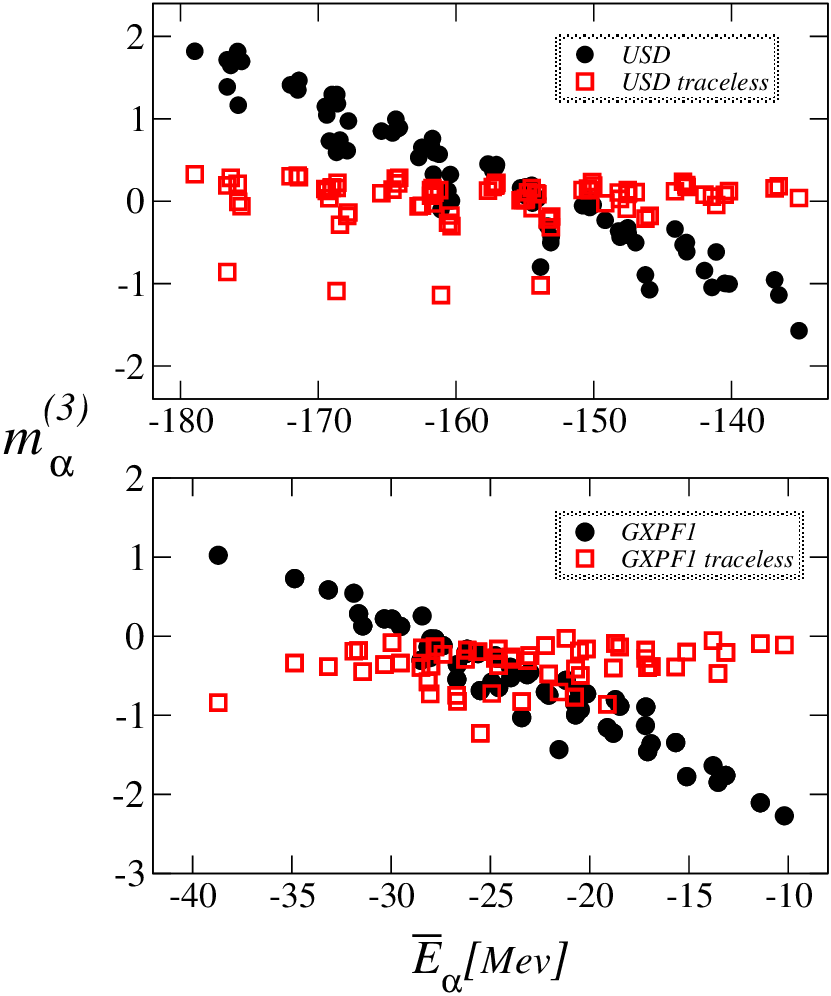}
\caption{\label{asymdist} (Color online) Configuration asymmetries
versus centroids for $^{33}$Ar (top) and $^{44}$Ti (bottom). Filled
circles are for the full interaction; open squares have the monopole
part of the interaction removed (``traceless''). }
\end{figure}

In Fig.~\ref{asymdist} we plot the configuration asymmetry against
the configuration centroid for two representative examples,
$^{33}$Ar in the $sd$-shell with the USD interaction (this is
similar to Fig.~1 of Ref.~\cite{KoPoSh86}, for 8 particles in the $sd$-shell,
also showing rather strange gill-like structure), and $^{44}$Ti in the
$pf$-shell using GXPF1.  Other nuclides in these spaces show the same generic
behavior.  The key results are, first, the \textit{range} of the
asymmetries, and second, the correlation of asymmetries with
centroids.  Please remember that here and throughout we consider scaled
asymmetries and excesses, that is, normalized by the relevant
configuration width.

The configuration asymmetries range from -2 to 2, which seems to be
a general limit; in other {\it sd}- and {\it pf}-shell nuclei, we
found only rare cases of $|m^{(3)}_\alpha|>2$  for $J$-summed moments
($J$-projected moments have wider ranges as we will see in a later
section). It is important to keep in mind that these findings are
only for complete $0\hbar \Omega$ spaces.  We have not yet
considered multi-shell configuration which may have a different
range; such spaces are beyond the scope of our current methods and
must be investigated in future.

The asymmetries are strongly correlated with the configuration
centroids $\bar{E}_\alpha$: the most positive asymmetries are found
at the most negative centroids and the the most negative asymmetries
at the most positive centroids.

Following the discussion in \ref{questions}, we recomputed the
asymmetries for ``traceless'' interactions (open squares) in
Fig~\ref{asymdist}, that is, we subtract off the monopole potential. (In such
a case the centroids all automatically go to zero, but we plot the
asymmetries against the original centroids so one may see how the
asymmetries are shifted.) Now most of the configuration asymmetries
go to nearly zero. This supports the hypothesis that much
of the configuration
asymmetry derives mostly, although not completely, 
from Eq.~(\ref{asym_cent}).

\begin{figure}[t]
\includegraphics[scale=0.65]{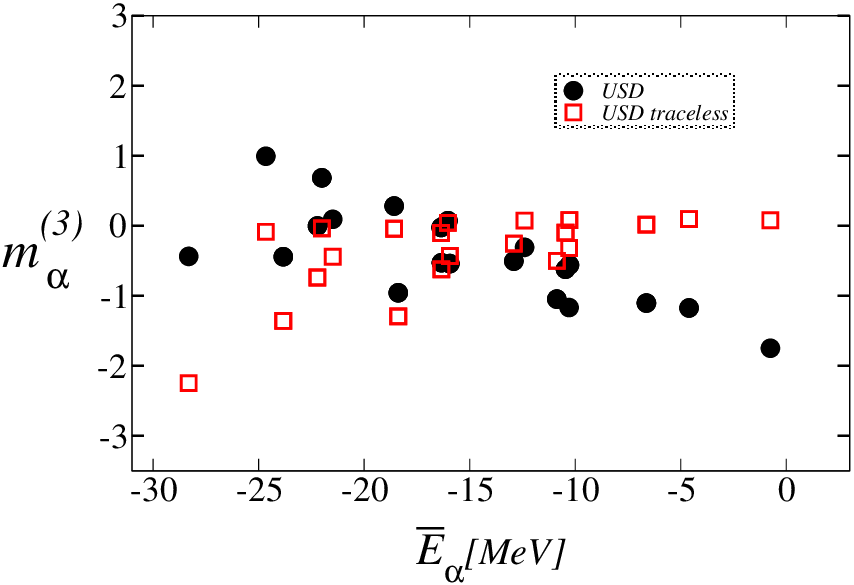}
\caption{\label{asymdist_Ne20} (Color online) Configuration asymmetries
versus centroids for $^{20}$Ne. Filled
circles are for the full interaction; open squares have the monopole
part of the interaction removed. }
\end{figure}
\begin{figure}
\includegraphics[scale=0.55]{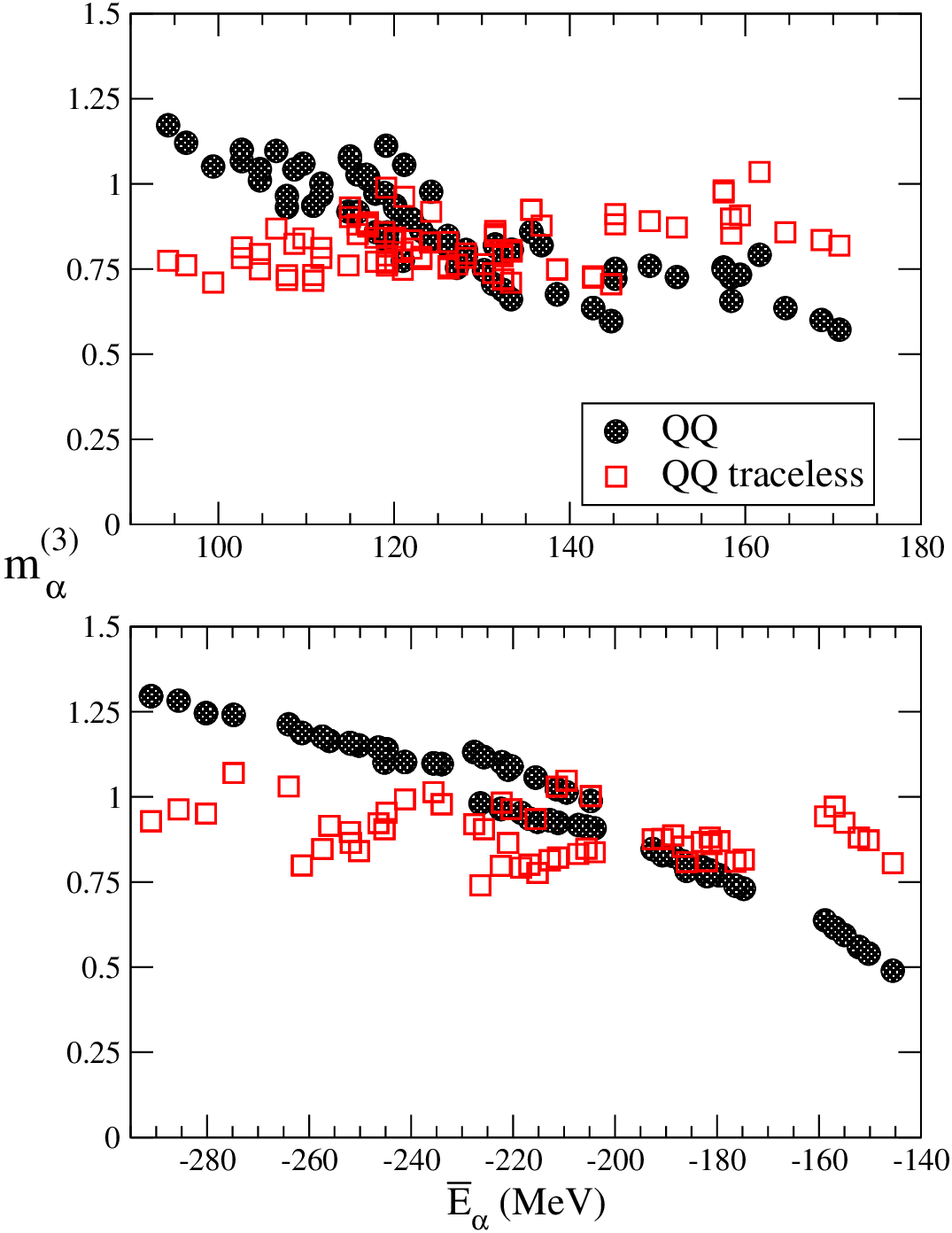}
\caption{\label{qqm3m4} (Color online) Configuration asymmetries
versus centroids for the collective $Q\cdot Q$ interactions for
$^{33}$Ar (top) and $^{44}$Ti (bottom). Filled
circles are for the full interaction; open squares have the monopole
part of the interaction removed. }
\end{figure}

We considered a number of nuclides and interactions and found consistently the
generic behavior illustrated in Fig.~\ref{asymdist}. There are two important
variations. First, we refer the reader to Fig.~1 of Ref~\cite{KoPoSh86}, which
for 8 particles in the $sd$ shell has a range of asymmetries from 0.5 at low
energies to -2 at high emergies. As another example, in
Fig.~\ref{asymdist_Ne20} we show the distribution of configuration asymmetries
for $^{20}$Ne, using the USD interaction. Several  of the asymmetries for
low-lying configuration, in particular the lowest energy configuration, have
asymmetries close to zero.  While these are important exceptions to the
examples in Fig.~\ref{asymdist}, the converse is also true: suggestions that
asymmetries are small at low energies \cite{KoPoSh86,HoKaZe03} do not
generalize.

For further investigation, in Fig.~\ref{asymdist_Ne20} we again consider the
asymmetries for a traceless USD interaction in $^{20}$Ne. Most of the asymmetries again go to
zero, but the low-energy asymmetries that were near zero now become large and
negative. We postulate that this arises from (probably attractive quadrupole)
collectivity in the lowest states, which would lead to a large negative
asymmetry. To test this idea we plot in Fig.~\ref{qqm3m4} asymmetries 
for a pure (repulsive) $Q\cdot Q$ interaction, both with and without (``traceless'')
the monopole-monopole interaction. In all our cases, including the two examples 
from the $sd$ shell ($^{33}$Ar, top) and $pf$ shell ($^{44}$Ti, bottom) 
we see strong asymmetries, even for the traceless interaction. 
In hindsight this is not surprising for a collective interaction,
but supports our interpretation of the behavior 
of more realistic interactions in Figs.~(\ref{asymdist}), (\ref{asymdist_Ne20}). 
A realistic interaction has both attractive and repulsive collective pieces,
e.g. $Q\cdot Q$ and $S \cdot S$, respectively, and in the bulk the collectivity 
tends to cancel, leading to  traceless asymmetries near zero. At low energy, 
however, the attractive quadrupole collectivity dominates, pushing the asymmetry 
lower. This effect is smaller in the $pf$ shell, where a larger spin-orbit 
splitting weakens quadrupole collectivity \cite{GuDrJo01}.  To summarize, the 
behavior of the configuration asymmetries and any correlation with the 
configuration centroids result from an interplay between collective interactions 
and the monopole-monopole interaction.

\subsection{Correlation of asymmetry and excess}

\begin{figure}[t]
\includegraphics[scale=0.63]{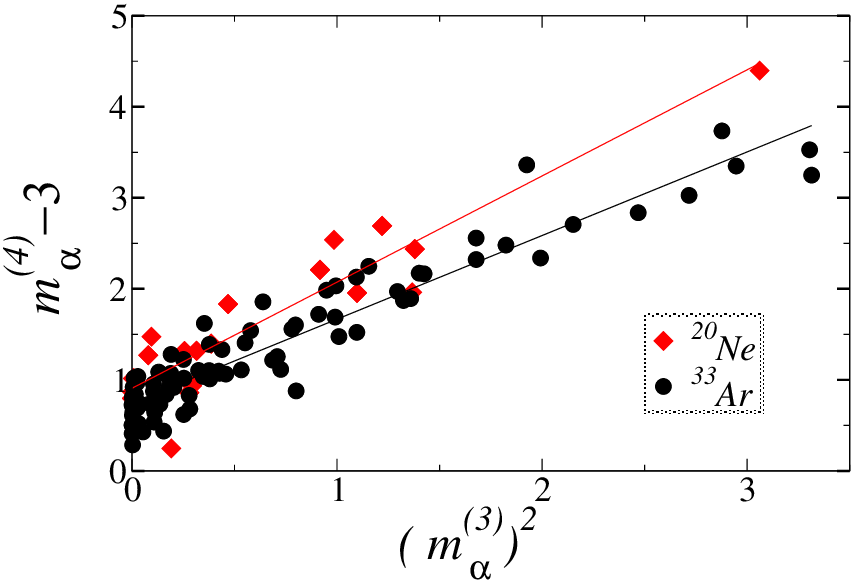}
\caption{(Color online) Correlation between configuration asymmetries and
excess in $sd$-shell nuclides $^{20}$Ne and $^{33}$Ar. Straight lines 
are least-squares linear fit. 
\label{m3m4correlation}}
\end{figure}

We now turn to look at the relationship between the third and the
fourth moments. Figure 1 of Ref.~\cite{KoPoSh86} indirectly suggest a
correlation but those authors did not comment on it. 
Fig. \ref{m3m4correlation} plots   $(m^{(3)}_\alpha)^2$ versus 
$m^{(4)}_\alpha$ for two nuclides in the $sd$-shell valence space, $^{20}$Ne and $^{33}$Ar.
The observed linear correlation occurs in all other nuclides we considered as well, and 
Table \ref{table:sdshell_linear} gives the
slope and intercept (as well as the corresponding ``uncertainties,'' which
are the diagonal correlation coefficients) for several
$sd$-shell nuclides, all with the USD interaction\cite{Wi84}.  Table
\ref{table:linearization} shows slope, intercept, and uncertainties for a
single nuclide in the $pf$-shell, $^{44}$Ti, but for a variety of
interactions, including
quadrupole-quadrupole and a member of the two-body random ensemble
(TBRE). There is a weak but statistically significant (the differences are
several standard deviations) dependence on
interaction and nuclide.

\begin{table}
 \caption{\label{table:sdshell_linear}
Intercept, slope,  and respective
 standard deviations ($\sigma_a$, $\sigma_b$) corresponding to linear fits of
 $m_4$ vs. $m_3^{2}$ plots for different nuclides in the $sd$-shell valence
space,
 with the USD interaction \cite{Wi84}.}
 \begin{ruledtabular}
 \begin{tabular}{ l c c c c}
 Nuclide       &   intercept    &  slope   & $\sigma_\mathrm{intercept}$  &
$\sigma_\mathrm{slope}$  \\
 \hline
%  $^{18}$F      &  0.25  & 1.08 &   0.550   &   0.202   \\
  $^{20}$Ne     &  0.90  & 1.16 &   0.066   &   0.080   \\
  $^{22}$Na     &  0.49  & 0.97 &   0.023   &   0.047   \\
  $^{34}$Cl     &  0.86  & 0.97 &   0.032   &   0.037   \\
  $^{33}$Ar     &  0.75  & 0.92 &   0.036   &   0.033   \\
 \end{tabular}
 \end{ruledtabular}
 \end{table}

 \begin{table}
 \caption{\label{table:linearization}
 Same as Table \ref{table:sdshell_linear} but for different
 interactions
on {\it pf}-shell $^{44}$Ti.}
 \begin{ruledtabular}
 \begin{tabular}{ l c c c c}
 Interaction    &   intercept    &  slope   & $\sigma_\mathrm{intercept}$  &
$\sigma_\mathrm{slope}$  \\
 \hline
  GXPF1         &  1.35  & 1.37 &   0.052   &   0.038   \\
  FPD6G         &  1.97  & 1.14 &   0.055   &   0.038   \\
  KB3G          &  2.30  & 1.17 &   0.090   &   0.032   \\
  $Q \cdot Q$         & -0.25  & 1.60 &   0.031   &   0.037   \\
  TBRE        & -0.70  & 1.88 &   0.029   &   0.085   \\
 \end{tabular}
 \end{ruledtabular}
 \end{table}

We attribute the relation between $m^{(3)}$ and $m^{(4)}$ to a maximum 
entropy principle. For convenience we took the distribution a modified 
Breit-Wigner,
\begin{equation}
\rho(E) = \rho_0 \frac{(E- E_\mathrm{min})(E_\mathrm{max} -E)}{ (E-E_0)^2 +W^2 },
\label{MBW}
\end{equation}
and can adjust the parameters to fix the moments. We chose the form 
(\ref{MBW}) because it is positive definite on the interval $(E_\mathrm{min}, E_\mathrm{max})$, 
and one can compute the moments analytically. (Some alternatives such as 
Gram-Charlier are not positive-definite; for yet other 
alternatives, such as Cornish-Fisher \cite{KoPoSh86} and binomial \cite{Zu01}, in 
practice one 
must integrate numerically to get \textit{accurate} moments at large asymmetries.) 
We then computed the entropy,
\begin{equation}
S = - \int_{E_\mathrm{min}}^{E_\mathrm{max}}\rho(E) \ln \rho(E) dE.
\end{equation}
For fixed width and normalization, the maximum entropy describes a straight line 
in the $m^{(3)})^2$-$m^{(4)}$ plane--albeit with different parameters, a slope of 1.67 and 
an intercept of $m^{(4)}-3 =-0.07$.  Nonetheless, this simple model makes for a reasonable understanding
of the correlation between the third and fourth moments.

\subsection{Distribution of $J$-projected moments}

Finally, we consider the behavior of $J$-projected moments. This is
relevant because some authors advocate the use of symmetric
(Gaussian) distributions for $J$-projected partial densities,
eschewing moments higher than the width
\cite{HaWo79,ZiBr81,HoKaZe03}. 

\begin{figure}
\includegraphics[scale=0.65]{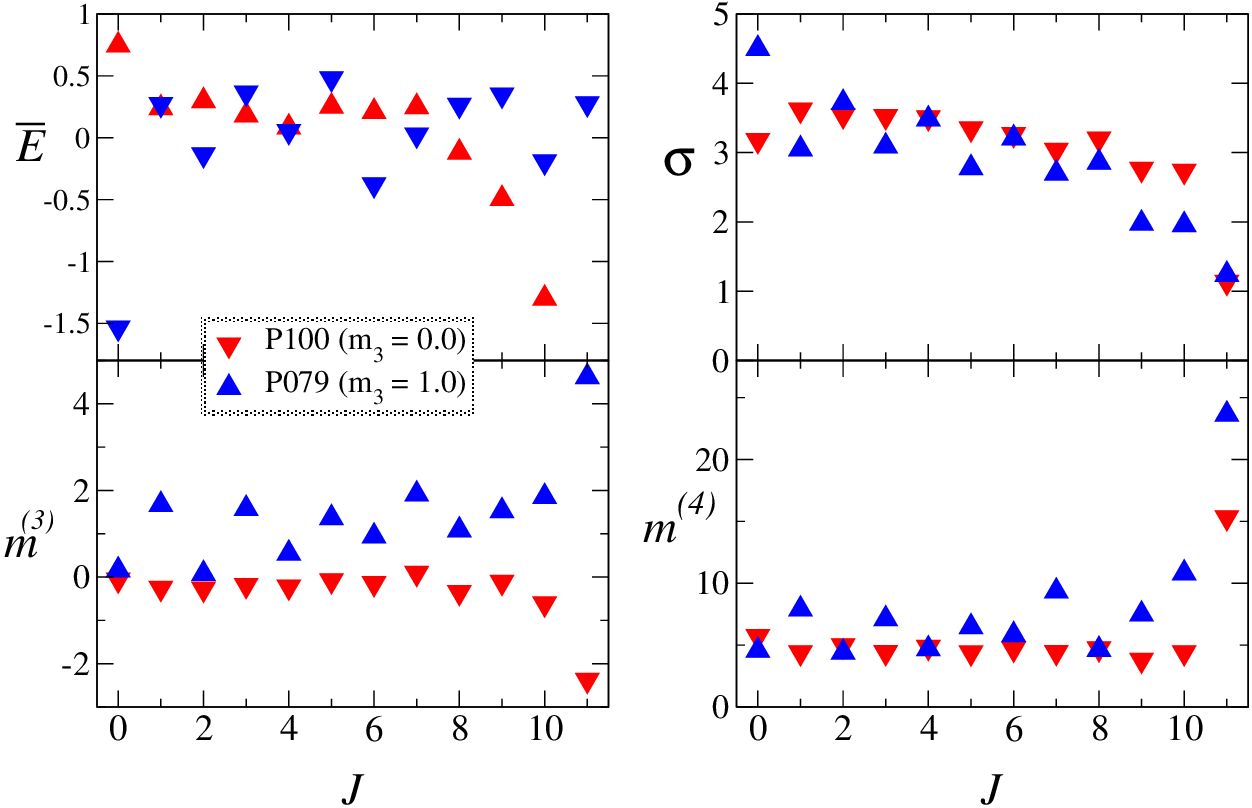}
\caption{\label{Jmom} (Color online)  Comparison of $J$-projected moments 
for $^{44}$Ti for two configurations.}
\end{figure}

In Fig.~\ref{Jmom} we plot $J$-projected moments for two representative
configurations for $^{44}$Ti using GXPF1. One
configuration (the 100th partition, or P100)  is nearly symmetric
with $m_3 = -0.12$, while  the other (P079) has high asymmetry ($m_3$ = 1.02).
The centroids are plotted relative to the corresponding $J$-summed
centroid, for ease of comparison.

The moments are relatively constant, although with larger scatter;  
this not surprising because of the smaller dimensions, and the 
reader is cautioned not to overinterpret these graphs. (For partition 100,
the total dimension is 784 levels, with 4 $J=0$ levels, 130 $J=6$ levels, 
and 25 $J=12$ levels; for partition 79, the total dimension is 1536 levels, 
4 $J=0$ levels, 234 $J=6$ levels, and 23 $J=11$ levels.)
The most significant 
trend is the width, which is largest for $J=0$ and shrinks almost
monotonically. This was found in previous investigations for $J$-projected
widths over the entire space\cite{VeBr82,GiYe75} and has been related to the
predominance of 
$J=0$ ground states even with random two-body
interactions\cite{BiFrPi99,PaWe04}. 
Similar behavior is found in other configurations in $^{44}$Ti as well 
as in other nuclides.

% In particular, only those partitions
%close to total symmetry, that is, with configuration centroid near the total
%centroid, have $J$-projected level densities lacking asymmetry.

\begin{figure}
\includegraphics[scale=0.65]{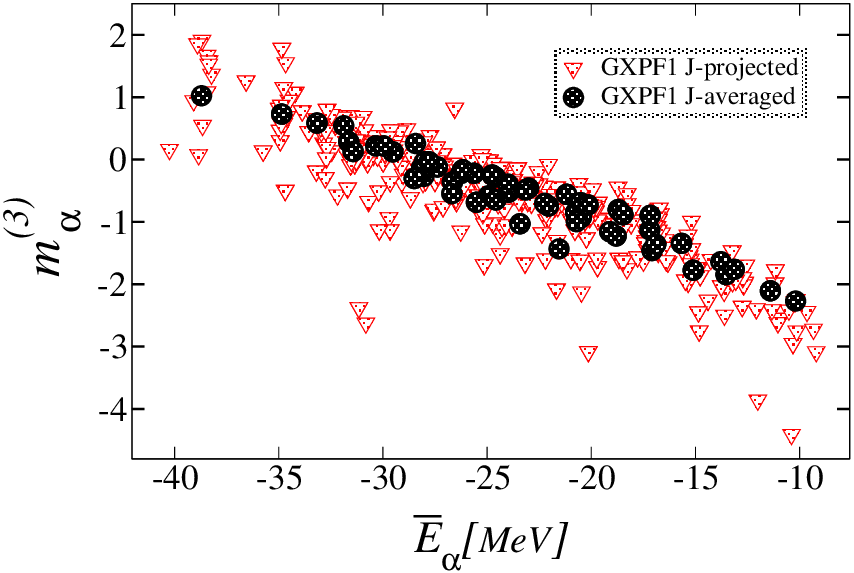}
\caption{\label{asymdist:jproj} (Color online) Asymmetries versus
centroids for $^{44}$Ti in the {\it pf}-shell with interaction GXPF1
\cite{HoOtBrMi04}. We compare $J$-summed asymmetries (filled
circles) against $J$-projected (triangles).}
\end{figure}

In Fig.~\ref{asymdist:jproj} we compare asymmetries versus centroids
for both $J$-summed and $J$-projected moments (again for $^{44}$Ti).
The $J$-projected moments track the $J$-summed moments.

\begin{figure}
\includegraphics[scale=0.65]{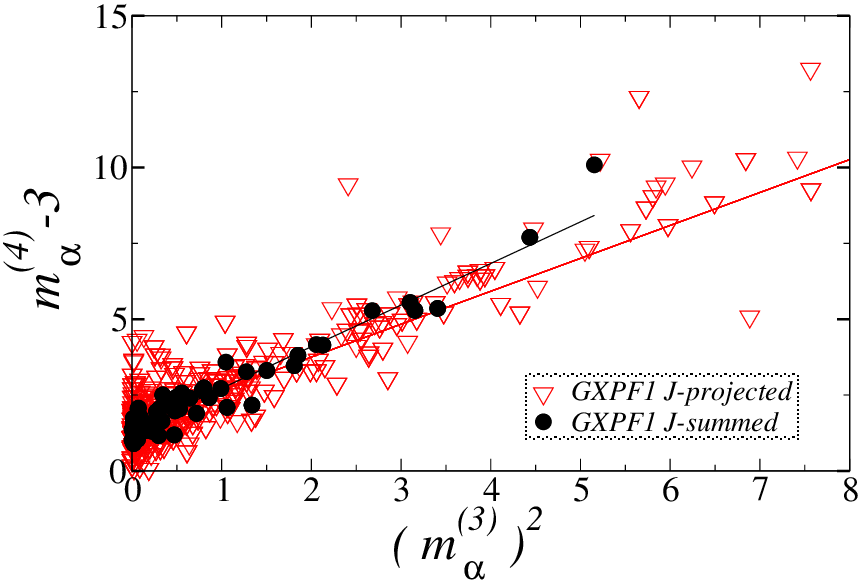}
\caption{ \label{m3m4:jproj} (Color online) Excess ($m_4$) plotted
against the square of the asymmetry ($m_3$) for $^{44}$Ti in the
{\it pf}-shell with interaction GXPF1 \cite{HoOtBrMi04}. We compare
$J$-summed asymmetries (filled circles) against $J$-projected
(triangles). Straight lines are least-squares linear fit. }
\end{figure}

Finally in Fig.~\ref{m3m4:jproj} we consider the correlation of
$m_4$ against $(m_3)^2$ for both $J$-projected and $J$-summed
moments (again for $^{44}$Ti). Both show the same linear relation,
although with slightly different slope and intercept. Also, the
$J$-projected moments have a greater range.

\section{Summary}

We have investigated in detail the distribution of and correlations
among configuration moments in $0\hbar\omega$ {\it sd}- and light {\it
pf}-shell
nuclei.  We regained earlier results finding:

\noindent $\bullet$ the configuration widths are constant (but
depend on the nuclide in question); 

\noindent $\bullet$ $J$-projected widths are largest for smallest $J$ and 
decrease as $J$ increases; and

\noindent $\bullet$ the configuration asymmetry (third moment) is
correlated with the configuration centroid (which, although documented before 
in the literature, had never been commented on).

We also found some new results, namely:

\noindent $\bullet$ the configuration asymmetries arise mostly, though not
entirely, through the monopole part of the interaction; and

\noindent $\bullet$ the configuration excesses are linearly
dependent upon the \textit{square} of the asymmetry, which we attribute 
to a maximum entropy principle.

These last two are particularly interesting. They suggest that, if
one wants to include the third and fourth moments, which are
computationally intensive, one might be able to take a
short cut and approximate the asymmetries from
Eqn.~(\ref{asym_cent})  and the fourth moments from the linear
relationship with the square of the asymmetry. Neither
``universality'' is absolute; intrinsic asymmetry (collectivity) at low energy
must be considered to get correct asymmetries, and the slope and intercept
of the $m^{(3)}$-$m^{(4)}$ relation depends weakly upon the interaction and
nuclide.  In the near future we plan to investigate in detail how
much error is induced through such approximations and indeed if
using the third and fourth moments do or do not improve descriptions
of the density of states.

There is one final consequence of our work: if one is indeed to
exploit knowledge of the asymmetry, any ``model function'' must be
able to handle a range of at least $-2 \leq m^{(3)}_\alpha \leq 2$. For
example, 
binomial distributions\cite{Zu01} fail to
work over this entire range. We will address this issue in 
future work.

\section{Acknowledgements}

This work is supported by grant DE-FG52-03NA00082  from the
Department of Energy /National Nuclear Security Agency.  CWJ 
acknowledges helpful conversations with Dr. W. E. Ormand and Dr. J.
Nabi.

%
%\newpage
\bibliography{biblio}

\begin{thebibliography}{36}
\expandafter\ifx\csname natexlab\endcsname\relax\def\natexlab#1{#1}\fi
\expandafter\ifx\csname bibnamefont\endcsname\relax
  \def\bibnamefont#1{#1}\fi
\expandafter\ifx\csname bibfnamefont\endcsname\relax
  \def\bibfnamefont#1{#1}\fi
\expandafter\ifx\csname citenamefont\endcsname\relax
  \def\citenamefont#1{#1}\fi
\expandafter\ifx\csname url\endcsname\relax
  \def\url#1{\texttt{#1}}\fi
\expandafter\ifx\csname urlprefix\endcsname\relax\def\urlprefix{URL }\fi
\providecommand{\bibinfo}[2]{#2}
\providecommand{\eprint}[2][]{\url{#2}}

\bibitem[{\citenamefont{Hauser and Feshbach}(1952)}]{HaFe52}
\bibinfo{author}{\bibfnamefont{W.}~\bibnamefont{Hauser}} \bibnamefont{and}
  \bibinfo{author}{\bibfnamefont{H.}~\bibnamefont{Feshbach}},
  \bibinfo{journal}{Phys. Rev.} \textbf{\bibinfo{volume}{87}},
  \bibinfo{pages}{366} (\bibinfo{year}{1952}).

\bibitem[{\citenamefont{{T. Rauscher, F.-K. Thielemann and K.-L.
  Kratz}}(1997)}]{RaThKr97}
\bibinfo{author}{\bibnamefont{{T. Rauscher, F.-K. Thielemann and K.-L.
  Kratz}}}, \bibinfo{journal}{Phys. Rev. C} \textbf{\bibinfo{volume}{56}},
  \bibinfo{pages}{1613} (\bibinfo{year}{1997}).

\bibitem[{\citenamefont{Bethe}(1936)}]{Be36}
\bibinfo{author}{\bibfnamefont{H.~A.} \bibnamefont{Bethe}},
  \bibinfo{journal}{Phys. Rev.} \textbf{\bibinfo{volume}{50}},
  \bibinfo{pages}{332} (\bibinfo{year}{1936}).

\bibitem[{\citenamefont{Goriely}(1996)}]{Go96}
\bibinfo{author}{\bibfnamefont{S.}~\bibnamefont{Goriely}},
  \bibinfo{journal}{Nucl. Phys. A} \textbf{\bibinfo{volume}{605}},
  \bibinfo{pages}{28} (\bibinfo{year}{1996}).

\bibitem[{\citenamefont{{P. Demetriou and S. Goriely}}(2001)}]{DeGo01}
\bibinfo{author}{\bibnamefont{{P. Demetriou and S. Goriely}}},
  \bibinfo{journal}{Nucl. Phys. A} \textbf{\bibinfo{volume}{695}},
  \bibinfo{pages}{95} (\bibinfo{year}{2001}).

\bibitem[{\citenamefont{{B.~A.~Brown, A.~Etchegoyen, and
  W.~D.~M.~Rae}}(1984)}]{OXBASH}
\bibinfo{author}{\bibnamefont{{B.~A.~Brown, A.~Etchegoyen, and W.~D.~M.~Rae}}}
  (\bibinfo{year}{1984}), \bibinfo{note}{{OXBASH, the Oxford University-Buenos
  Aires-MSU shell model code, Michigan State University Cyclotron Laboratory
  Report No. 524}}.

\bibitem[{\citenamefont{{E. Caurier and F. Nowacki}}(1999)}]{ANTOINE}
\bibinfo{author}{\bibnamefont{{E. Caurier and F. Nowacki}}},
  \bibinfo{journal}{Acta Phys. Pol. B} \textbf{\bibinfo{volume}{30}},
  \bibinfo{pages}{705} (\bibinfo{year}{1999}).

\bibitem[{\citenamefont{Ormand}()}]{Redstick}
\bibinfo{author}{\bibfnamefont{W.~E.} \bibnamefont{Ormand}},
  \bibinfo{note}{{REDSTICK shell model code, private communication}}.

\bibitem[{\citenamefont{{R.R. Whitehead, A.~ Watt, B.J.~Cole, and I.
  Morrison}}(1977)}]{Wh77}
\bibinfo{author}{\bibnamefont{{R.R. Whitehead, A.~ Watt, B.J.~Cole, and I.
  Morrison}}}, \bibinfo{journal}{Adv. Nucl.~Phys.}
  \textbf{\bibinfo{volume}{9}}, \bibinfo{pages}{123} (\bibinfo{year}{1977}).

\bibitem[{\citenamefont{Ormand}(1997)}]{Or97}
\bibinfo{author}{\bibfnamefont{W.~E.} \bibnamefont{Ormand}},
  \bibinfo{journal}{Phys. Rev. C} \textbf{\bibinfo{volume}{56}},
  \bibinfo{pages}{R1678} (\bibinfo{year}{1997}).

\bibitem[{\citenamefont{Nakada and Alhassid}(1997)}]{NaAl97}
\bibinfo{author}{\bibfnamefont{H.}~\bibnamefont{Nakada}} \bibnamefont{and}
  \bibinfo{author}{\bibfnamefont{Y.}~\bibnamefont{Alhassid}},
  \bibinfo{journal}{Phys. Rev. Lett.} \textbf{\bibinfo{volume}{79}},
  \bibinfo{pages}{2939} (\bibinfo{year}{1997}).

\bibitem[{\citenamefont{Nakada and Alhassid}(1998)}]{NaAl98}
\bibinfo{author}{\bibfnamefont{H.}~\bibnamefont{Nakada}} \bibnamefont{and}
  \bibinfo{author}{\bibfnamefont{Y.}~\bibnamefont{Alhassid}},
  \bibinfo{journal}{Phys. Lett. B} \textbf{\bibinfo{volume}{436}},
  \bibinfo{pages}{231} (\bibinfo{year}{1998}).

\bibitem[{\citenamefont{{J. B. French and K. F. Ratcliff}}(1971)}]{FrRa71}
\bibinfo{author}{\bibnamefont{{J. B. French and K. F. Ratcliff}}},
  \bibinfo{journal}{Phys. Rev. C} \textbf{\bibinfo{volume}{3}},
  \bibinfo{pages}{94} (\bibinfo{year}{1971}).

\bibitem[{\citenamefont{{S. Ayik and J.N. Ginocchio}}(1974)}]{AyGi74}
\bibinfo{author}{\bibnamefont{{S. Ayik and J.N. Ginocchio}}},
  \bibinfo{journal}{Nucl. Phys. A} \textbf{\bibinfo{volume}{221}},
  \bibinfo{pages}{285} (\bibinfo{year}{1974}).

\bibitem[{\citenamefont{{K. K. Mon and J. B. French}}(1975)}]{MoFr75}
\bibinfo{author}{\bibnamefont{{K. K. Mon and J. B. French}}},
  \bibinfo{journal}{Ann. Phys.} \textbf{\bibinfo{volume}{95}},
  \bibinfo{pages}{90} (\bibinfo{year}{1975}).

\bibitem[{\citenamefont{Wong}(1986)}]{Wo86}
\bibinfo{editor}{\bibfnamefont{S.~S.~M.} \bibnamefont{Wong}}, ed.,
  \emph{\bibinfo{title}{Nuclear Statistical Spectroscopy}}
  (\bibinfo{publisher}{Oxford University Press}, \bibinfo{address}{New York},
  \bibinfo{year}{1986}).

\bibitem[{\citenamefont{Kota}(1984)}]{Ko84}
\bibinfo{author}{\bibfnamefont{V.~K.~B.} \bibnamefont{Kota}},
  \bibinfo{journal}{Z. Phys. A} \textbf{\bibinfo{volume}{315}},
  \bibinfo{pages}{91} (\bibinfo{year}{1984}).

\bibitem[{\citenamefont{{V. K. B. Kota, V. Potbhare and P.
  Shenoy}}(1986)}]{KoPoSh86}
\bibinfo{author}{\bibnamefont{{V. K. B. Kota, V. Potbhare and P. Shenoy}}},
  \bibinfo{journal}{Phys. Rev. C} \textbf{\bibinfo{volume}{34}},
  \bibinfo{pages}{2330} (\bibinfo{year}{1986}).

\bibitem[{\citenamefont{Zuker}(2001)}]{Zu01}
\bibinfo{author}{\bibfnamefont{A.~P.} \bibnamefont{Zuker}},
  \bibinfo{journal}{Phys. Rev. C} \textbf{\bibinfo{volume}{64}},
  \bibinfo{pages}{021303} (\bibinfo{year}{2001}).

\bibitem[{\citenamefont{Horoi et~al.}(2003)\citenamefont{Horoi, Kaiser, and
  Zelevinsky}}]{HoKaZe03}
\bibinfo{author}{\bibfnamefont{M.}~\bibnamefont{Horoi}},
  \bibinfo{author}{\bibfnamefont{J.}~\bibnamefont{Kaiser}}, \bibnamefont{and}
  \bibinfo{author}{\bibfnamefont{V.}~\bibnamefont{Zelevinsky}},
  \bibinfo{journal}{Phys. Rev. C} \textbf{\bibinfo{volume}{67}},
  \bibinfo{pages}{054309} (\bibinfo{year}{2003}).

\bibitem[{\citenamefont{Duflo and Zuker}(1999)}]{DuZu99}
\bibinfo{author}{\bibfnamefont{J.}~\bibnamefont{Duflo}} \bibnamefont{and}
  \bibinfo{author}{\bibfnamefont{A.~P.} \bibnamefont{Zuker}},
  \bibinfo{journal}{Phys. Rev. C} \textbf{\bibinfo{volume}{59}},
  \bibinfo{pages}{R2347} (\bibinfo{year}{1999}).

\bibitem[{\citenamefont{Teran and Johnson}()}]{TeJo05}
\bibinfo{author}{\bibfnamefont{E.}~\bibnamefont{Teran}} \bibnamefont{and}
  \bibinfo{author}{\bibfnamefont{C.~W.} \bibnamefont{Johnson}},
  \bibinfo{note}{to be published}.

\bibitem[{\citenamefont{Wildenthal}(1984)}]{Wi84}
\bibinfo{author}{\bibfnamefont{B.~H.} \bibnamefont{Wildenthal}},
  \bibinfo{journal}{Prog. Part. Nucl. Phys.} \textbf{\bibinfo{volume}{11}},
  \bibinfo{pages}{5} (\bibinfo{year}{1984}).

\bibitem[{\citenamefont{Honma et~al.}(2004)\citenamefont{Honma, Otsuka, Brown,
  and Mizusaki}}]{HoOtBrMi04}
\bibinfo{author}{\bibfnamefont{M.}~\bibnamefont{Honma}},
  \bibinfo{author}{\bibfnamefont{T.}~\bibnamefont{Otsuka}},
  \bibinfo{author}{\bibfnamefont{B.~A.} \bibnamefont{Brown}}, \bibnamefont{and}
  \bibinfo{author}{\bibfnamefont{T.}~\bibnamefont{Mizusaki}},
  \bibinfo{journal}{Physical Review C} \textbf{\bibinfo{volume}{69}},
  \bibinfo{eid}{034335} (\bibinfo{year}{2004}).

\bibitem[{\citenamefont{{W. A. Richter, M. J. Van Der Merwe, R. R. Julies and
  B. A. Brown}}(1991)}]{RMJB}
\bibinfo{author}{\bibnamefont{{W. A. Richter, M. J. Van Der Merwe, R. R. Julies
  and B. A. Brown}}}, \bibinfo{journal}{Nucl. Phys. A}
  \textbf{\bibinfo{volume}{523}}, \bibinfo{pages}{325} (\bibinfo{year}{1991}).

\bibitem[{\citenamefont{{A. Poves, J. S\'anchez-Solano, E. Caurier and F.
  Nowacki}}(2001)}]{PoSaCaNo01}
\bibinfo{author}{\bibnamefont{{A. Poves, J. S\'anchez-Solano, E. Caurier and F.
  Nowacki}}}, \bibinfo{journal}{Nucl. Phys. A} \textbf{\bibinfo{volume}{694}},
  \bibinfo{pages}{157} (\bibinfo{year}{2001}).

\bibitem[{\citenamefont{{T.T.S.~Kuo and G.E.~Brown}}(1968)}]{Kubr68}
\bibinfo{author}{\bibnamefont{{T.T.S.~Kuo and G.E.~Brown}}},
  \bibinfo{journal}{Nucl. Phys. A} \textbf{\bibinfo{volume}{114}},
  \bibinfo{pages}{241} (\bibinfo{year}{1968}).

\bibitem[{\citenamefont{{A.~Poves and A.P.~Zuker}}(1981)}]{PoZu81}
\bibinfo{author}{\bibnamefont{{A.~Poves and A.P.~Zuker}}},
  \bibinfo{journal}{Phys. Rep.} \textbf{\bibinfo{volume}{70}},
  \bibinfo{pages}{235} (\bibinfo{year}{1981}).

\bibitem[{\citenamefont{{C. Jacquemin and S. Spitz}}(1979)}]{JaSp79}
\bibinfo{author}{\bibnamefont{{C. Jacquemin and S. Spitz}}},
  \bibinfo{journal}{J. Phys. G} \textbf{\bibinfo{volume}{5}},
  \bibinfo{pages}{L95} (\bibinfo{year}{1979}).

\bibitem[{\citenamefont{{J.J.M. Verbaarschot and P.J.
  Brussaard}}(1982)}]{VeBr82}
\bibinfo{author}{\bibnamefont{{J.J.M. Verbaarschot and P.J. Brussaard}}},
  \bibinfo{journal}{Phys. Lett. B} \textbf{\bibinfo{volume}{102}},
  \bibinfo{pages}{201} (\bibinfo{year}{1982}).

\bibitem[{\citenamefont{{V. G. Gueorguiev, J. P. Draayer, and C. W.
  Johnson}}(2001)}]{GuDrJo01}
\bibinfo{author}{\bibnamefont{{V. G. Gueorguiev, J. P. Draayer, and C. W.
  Johnson}}}, \bibinfo{journal}{Phys. Rev. C} \textbf{\bibinfo{volume}{63}},
  \bibinfo{pages}{014318} (\bibinfo{year}{2001}).

\bibitem[{\citenamefont{{R. U. Haq and S. S. M. Wong}}(1979)}]{HaWo79}
\bibinfo{author}{\bibnamefont{{R. U. Haq and S. S. M. Wong}}},
  \bibinfo{journal}{Nucl. Phys. A} \textbf{\bibinfo{volume}{327}},
  \bibinfo{pages}{314} (\bibinfo{year}{1979}).

\bibitem[{\citenamefont{{M. R. Zirnbauer and D. M. Brink}}(1981)}]{ZiBr81}
\bibinfo{author}{\bibnamefont{{M. R. Zirnbauer and D. M. Brink}}},
  \bibinfo{journal}{Z. Phys. A} \textbf{\bibinfo{volume}{301}},
  \bibinfo{pages}{237} (\bibinfo{year}{1981}).

\bibitem[{\citenamefont{{J.N. Ginocchio and M.M. Yen}}(1975)}]{GiYe75}
\bibinfo{author}{\bibnamefont{{J.N. Ginocchio and M.M. Yen}}},
  \bibinfo{journal}{Nucl. Phys. A} \textbf{\bibinfo{volume}{239}},
  \bibinfo{pages}{365} (\bibinfo{year}{1975}).

\bibitem[{\citenamefont{{ R. Bijker, A. Frank, and S.
  Pittel}}(1999)}]{BiFrPi99}
\bibinfo{author}{\bibnamefont{{ R. Bijker, A. Frank, and S. Pittel}}},
  \bibinfo{journal}{Phys. Rev. C} \textbf{\bibinfo{volume}{60}},
  \bibinfo{pages}{021302} (\bibinfo{year}{1999}).

\bibitem[{\citenamefont{{ T. Papenbrock and H. A.
  Weidenm\"uller}}(2004)}]{PaWe04}
\bibinfo{author}{\bibnamefont{{ T. Papenbrock and H. A. Weidenm\"uller}}},
  \bibinfo{journal}{Phys. Rev. Lett.} \textbf{\bibinfo{volume}{93}},
  \bibinfo{pages}{132503} (\bibinfo{year}{2004}).

\end{thebibliography}

\end{document}